\documentclass{ws-procs10x7}

\begin{document}

\title{Supersymmetry at the Linear Collider}

\author{SEONG YOUL CHOI}

\address{Department of Physics, Chonbuk National University,
         Chonju 561--756, Korea\\ E-mail: sychoi@chonbuk.ac.kr}

\twocolumn[\maketitle\abstract{If supersymmetry (SUSY) is realized at the
          electroweak scale, its underlying structure and breaking mechanism may
          be explored with great precision by a future linear $e^+ e^-$ collider
          (LC) with a clean environment, tunable collision energy, high luminosity
          polarized beams, and additional $e^-e^-$, $e\gamma$ and $\gamma\gamma$
          modes. In this report we summarize four papers submitted
          to the ICHEP04 conference about the precise measurements of the top
          squark parameters and $\tan\beta$, the impacts of the CP phases on the
          search for top/bottom squarks, the Majorana nature and CP violation
          in the neutralino system, the implications of the SUSY dark matter
          scenario for the LC experiments, and the characteristics of
          the neutralino sector of the next--to--minimal supersymmetric standard
          model at the LC.}]

\section{Introduction}
\label{sec:introduction}

Weak--scale SUSY has its natural solution to the gauge hierarchy
problem, providing a {\it stable} bridge between the electroweak scale
and the grand unification or Planck scale\cite{stable}, with which
the roots of standard particle physics are expected to go as deep as the Planck
length of $10^{-33}$\,cm. It is then crucial to probe SUSY and its
breaking with great precision at a future $e^+e^-$ linear collider
(LC)\cite{LC} as well as the large hadron collider (LHC)\cite{LHC}
for a reliable grand extrapolation to the Planck scale\cite{RGE}.

In this report we summarize four papers submitted to the ICHEP04
conference about SUSY phenomenology at the LC.

\section{Precise determinations of SUSY parameters}
\label{sec:precise}

If SUSY is realized at the electroweak scale, the LC experiments can be
performed in the SUSY sector with high precision. In this section, I report on
two related works submitted to this conference.

\subsection{Top squark mass determinations}

The study of the top squarks is of particular interest, since the lighter
top squark is likely to be the lightest squark in a SUSY theory due to the
significant mixing between two top squark weak eigenstates.

The recent work in Ref.~\cite{stop_parameter} compares four methods for
measuring the top squark mass $m_{\tilde{t}_1}$ at the LC. Two conventional
methods rely on an accurate measurement of the production cross section with
beam polarization at one fixed energy and through threshold scans of the cross
section\footnote{The first method allows us to measure the mixing angle
$\theta_{\tilde{t}}$ as well as the top squark mass.}. The other two methods for
measuring the top squark
mass use information from two measured charm jets with large missing energy
due to the unobserved neutralino $\tilde{\chi}^0_1$.
\begin{figure}[htb]
\begin{center}
\includegraphics*[width=3.cm,height=3.3cm]{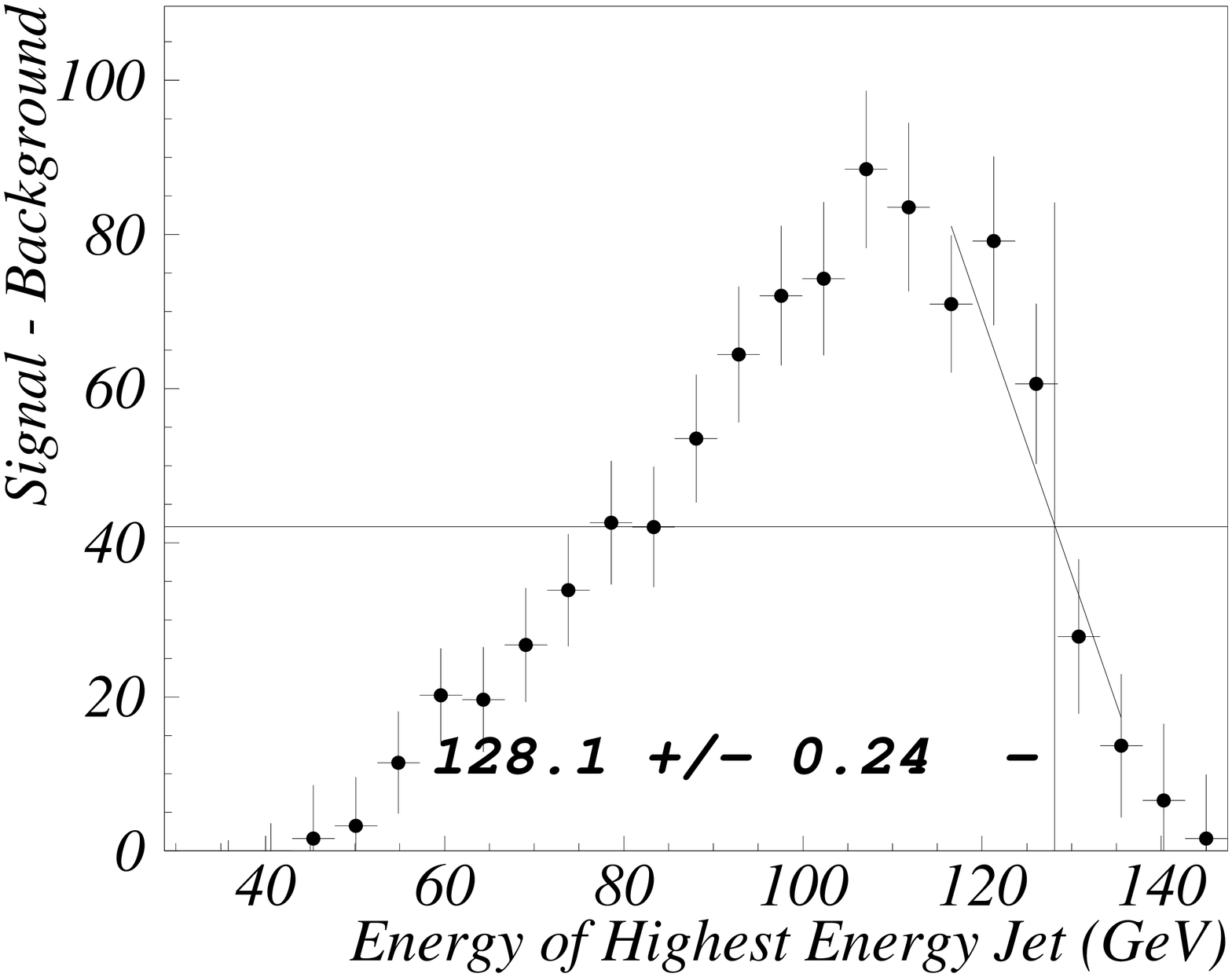}
\hskip 0.4cm
\includegraphics*[width=3.cm,height=3.3cm]{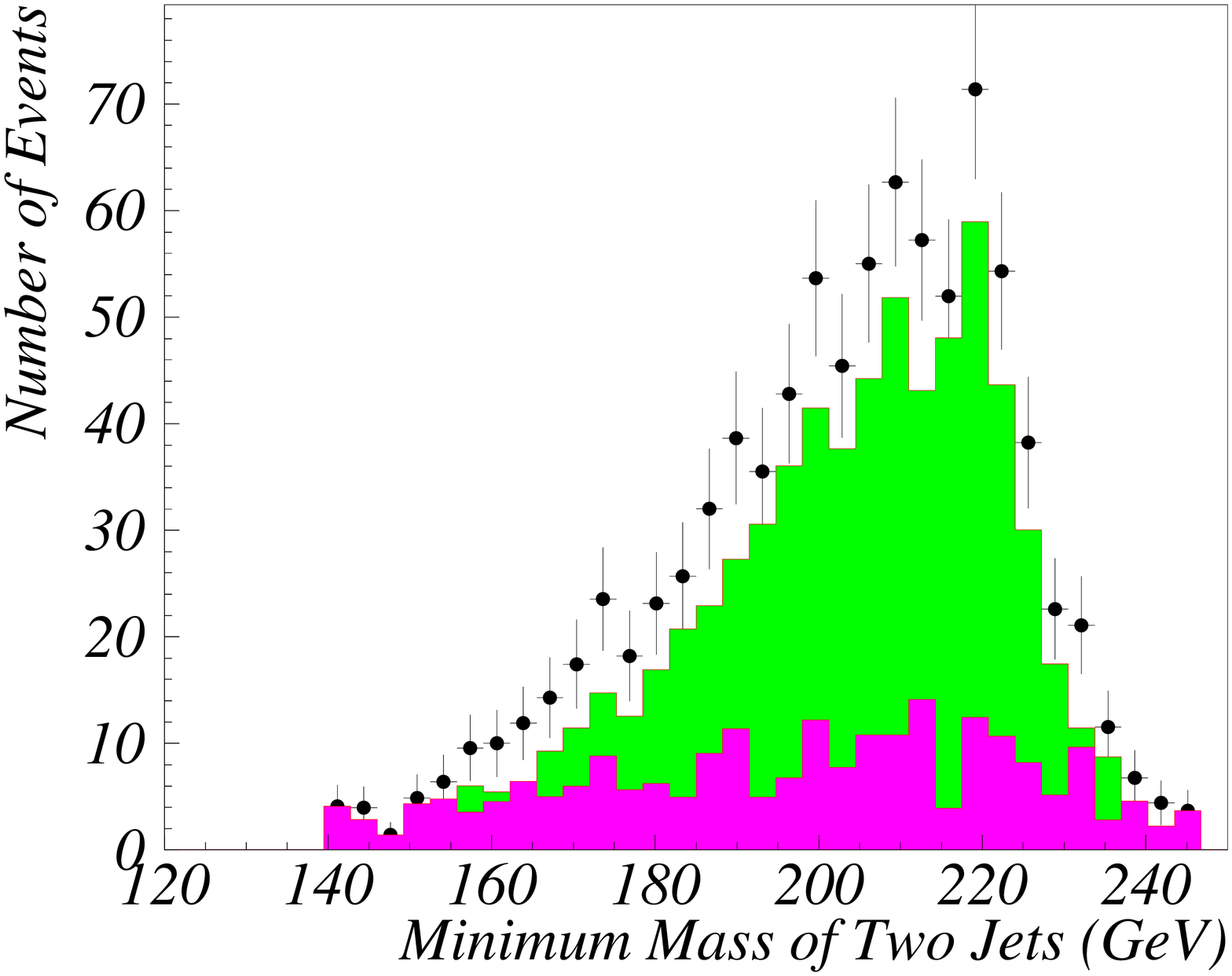}
\caption{Examples of measuring the maximum jet energy end point (left) and
         the minimum mass of two jets (right).}
     \label{fig:stop_mass}
\end{center}
\end{figure}

Two end points of the charm energy spectrum of
$\tilde{t}_1\to c\tilde{\chi}^0_1$, flat at the parton level,
contain information on $m_{\tilde{t}_1}$ and $m_{\tilde{\chi}_1}$.
In practice, this ideal situation is distorted by resolution effects as
demonstrated in the left panel of Fig.~\ref{fig:stop_mass} for the SPS5
point\cite{SPS}. When $m_{\tilde{\chi}^0_1}$ is known the
minimum allowed mass distribution of the two charm jets in an event
can be used to measure $m_{\tilde{t}_1}$ with the accuracy of
the order of 1 GeV with $\int {\cal L} dt = 500$ fb$^{-1}$ at
$\sqrt{s}=500$ GeV, comparable to that from the other methods,
as shown in the right panel of Fig.~\ref{fig:stop_mass}.

\subsection{A new $\tan\beta$ determination method}

Many observables, in the chargino/neutralino
sector\cite{chi_sector,REC} for instance, involve only $\cos
2\beta$ and thus are quite insensitive to $\tan\beta$ for large
values. On the contrary, for large pseudoscalar Higgs  mass the
heavy $H/A$ Higgs couplings to down-type fermions are directly
proportional to $\tan\beta$ if the parameter is large so that they
are highly sensitive to its value\cite{R2}. Also the down--type
couplings of the light $h$ Higgs boson in the MSSM are close to
$\tan\beta$ if $M_A$ is moderately small. Based on these
observations, we show that $\tau\tau$ fusion to Higgs bosons at a
photon collider\cite{2a} can provide a valuable method for
measuring $\tan\beta$.

For large $\tan\beta$, all the Higgs bosons $\Phi$ ($=H, A, h$)
decay almost exclusively [80 to 90\%] to a pair of $b$ quarks so
that the final state consists of a pair of $\tau$'s and a pair of
resonant $b$ quark jets. Two main background processes -- the
$\tau^+ \tau^-$ annihilation into a pair of $b$-quarks via
$s$--channel $\gamma/Z$ exchanges and the diffractive $\gamma
\gamma \to (\tau^+ \tau^-) (b\bar{b})$ events with the pairs
scattering off each other by Rutherford photon exchange -- can be
suppressed strongly by choosing proper cuts\cite{2a}.

\begin{figure}[!htb]
\begin{center}
\includegraphics*[width=3cm, height=3.5cm]{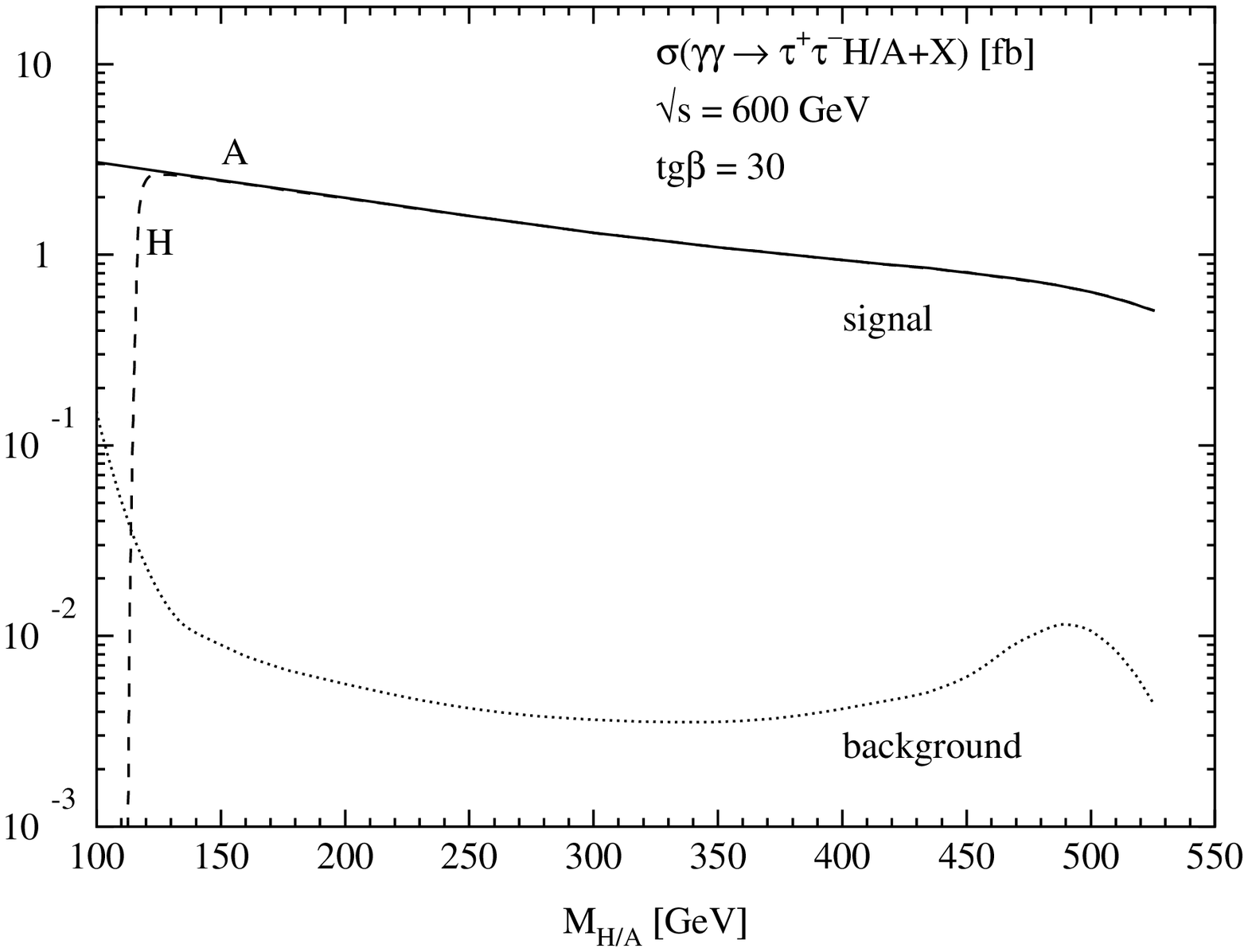}
\hskip 0.4cm
\includegraphics*[width=3cm, height=3.5cm]{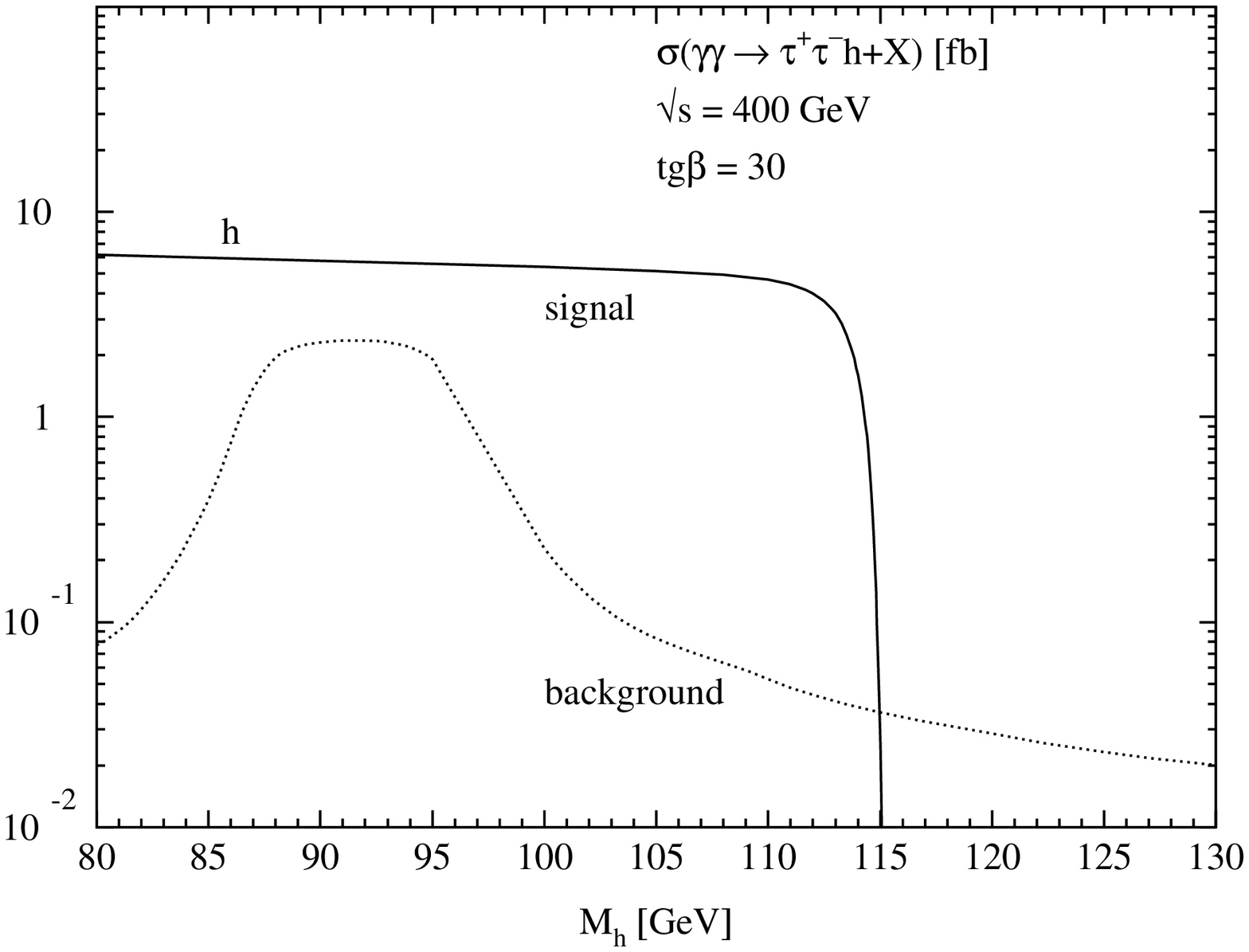}
\caption{The cross sections for the production of the $H/A$
  (left) and $h$ (right) Higgs bosons in the $\tau\tau$ fusion process
  at a $\gamma\gamma$ collider for $\tan\beta=30$.  Also shown is the
  background cross section with experimental cuts. $\sqrt{s}$ denotes
  the $\gamma\gamma$ collider c.m. energy.}
\label{fig:xsec}
\end{center}
\end{figure}

The left panel of Fig.~\ref{fig:xsec} shows the exact cross
sections for the signals of $H$ and $A$ Higgs--boson production in
the $\tau\tau$ fusion process with $E_{\gamma\gamma}=600$ GeV,
together with all the background processes with appropriate
experimental cuts. As shown in the right panel of
Fig.~\ref{fig:xsec} $\tau\tau$ fusion to the light Higgs boson $h$
with $E_{\gamma\gamma}=400$ GeV can also be exploited to measure
large $\tan\beta$ for moderately small $M_A$. For $h$ production,
the mass parameters are set to $M_A\sim 100$ GeV and $M_h=100$
GeV. The channels $h/A$ and $H/A$ are combined in the overlapping
mass ranges in which the respective two states cannot be
discriminated. Since in the region of interest the $\tau\tau$
fusion cross sections are proportional to $\tan^2\beta$ and the
background is small, {\it the absolute errors $\Delta\tan\beta$ are
nearly independent of $\tan\beta$}, varying between $\sim 0.9$ and
$1.3$ for Higgs masses away from the kinematical limits for the
integrated luminosity of 200/100 fb$^{-1}$ for the high/low energy
option.

\section{CP violation in the MSSM}
\label{sec:cp_violation}

Many SUSY parameters in the MSSM are in general complex,
in particular the higgsino mass parameter $\mu$, the gaugino mass parameters
$M_{1,2,3}$ and the trilinear scalar coupling parameters $A_f$ of the
sfermions $\tilde{f}$\footnote{The SU(2) gaugino mass parameter $M_2$ can be set
real and positive after an appropriate redefinition of the fields.}.

Not only the CP--violating observables such as electric dipole moments\cite{EDM}
and triple products of momenta and polarization vectors\cite{triple} but also the
CP--conserving observables like cross sections and decay widths depend
on the phases of the complex parameters. Recently there have been a lot of
interesting works on the direct and indirect observations of CP violation in
the SUSY particle sectors\cite{EDM,triple,sparticle_CP} and the
Higgs boson sector\cite{Higgs_CP,Higgs_CP1} of the CP--noninvariant version of
the MSSM. In this section we review two relavant works submitted to this
ICHEP04 conference.

\subsection{Impacts of CP phases on the top/bottom squark searches}

 The phases of
the trilinear parameter $A_f$ and the higgsino mass parameter $\mu$ are
involved directly in the squark mass matrices and the squark--Higgs
and squark--quark--gaugino/Higgsino couplings. As a result, the
squark--pair production cross sections and squark decay widths are
strongly affected by the phases. In Ref.~\cite{third_CP} the authors
have studied the effects of the phases of the parameters $A_{t,b}$, $\mu$
and $M_1$ on the phenomenology of the top/bottom squarks, which can be
significant due to large Yukawa couplings.

\begin{figure}[htb]
\begin{center}
\includegraphics*[width=6.cm,height=3.3cm]{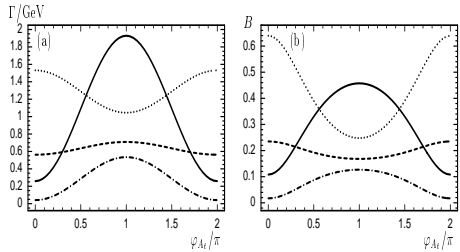}
\caption{The $\phi_{A_t}$ dependence of the partial decay widths (left) and
         branching ratios (right) for the decays $\tilde{t}_1\to\tilde{\chi}^+_1
         b$ (solid), $\tilde{t}_1\to\tilde{\chi}^0_1 t$ (dashed),
         $\tilde{t}_1\to \tilde{\chi}^+_2 b$ (dash--dotted) and
         $\tilde{t}_1\to\tilde{\chi}^0_2\,t$ (dotted) for the parameter set
         described in the text.
         }
\label{fig:stop_cp}
\end{center}
\end{figure}

As clearly shown in Fig.~\ref{fig:stop_cp} with a typical parameter set,
the partial widths (left) and branching ratios (right) for the top
squark decays depend strongly on the CP phase $\phi_{A_t}$, implying the
importance of taking into account the impacts of the CP phases in searching
for SUSY particles.

\subsection{Majorana nature and CP violation in the neutralino system}

It is a unique SUSY test to establish the Majorana nature and
CP properties of neutralinos. Here, we describe two
methods for probing the Majorana nature and CP violation in the
neutralino system.

With the neglected SM fermion masses both the processes, $e^+e^- \to
\tilde{\chi}^0_i\tilde{\chi}^0_j$ and $\tilde{\chi}^0_i \to
\tilde{\chi}^0_j f\bar{f}$, can effectively be regarded as
processes of a vector current exchange between two neutralinos.
In the CP invariant case, the neutralino $\{ij\}$ pair production and the
decay $\tilde{\chi}^0_i \to \tilde{\chi}^0_j\, V$ through a vector
current satisfy the CP relations
\begin{eqnarray}
1\, =\, \pm\eta^i \eta^j\, \left(-1\right)^L \label{eq:selection}
\end{eqnarray}
for static neutralinos, with $\eta^i=\pm i$ the intrinsic
$\tilde{\chi}^0_i$ CP parity and $L$ the orbital angular momentum
of the produced pair $\{ij\}$ and of the final state of
$\tilde{\chi}^0_j$ and $V$, respectively. Therefore, in the CP
invariant case, if the production of a pair of neutralinos with
the same (opposite) CP parity is excited slowly in $P$ waves
(steeply in $S$ waves) near threshold, then the $\tilde{\chi}^0_i$
to $\tilde{\chi}^0_j$ transition is excited sharply in $S$ waves
(slowly in $P$ waves) near the end point of the fermion invariant
mass.

In the CP noninvariant case the orbital angular momentum is no
longer restricted by the selection rules (\ref{eq:selection}).
Consequently, CP violation in the neutralino system can  be
signalled by (a) the sharp $S$--wave excitations of the production
of three non--diagonal $\{ij\}$, $\{ik\}$ and $\{jk\}$ pairs near
threshold\cite{REC,Threshold1} or by (b) the simultaneous
$S$--wave excitations of the production of any non--diagonal
$\{ij\}$ pair in $e^+e^-$ annihilation near threshold and of the
fermion invariant mass distribution of the neutralino three--body
decays $\tilde{\chi}^0_i \to \tilde{\chi}^0_j f\bar{f}$ near the
kinematical end point\cite{Threshold2}. Note that even the
combined analysis of the lighter neutralino
$\{12\}$ pair production and the associated decay $\tilde{\chi}^0_2 \to
\tilde{\chi}^0_1 f\bar{f}$ enables us to probe CP violation in the
neutralino system.

In addition, if the two--body decays $\tilde{\chi}^0_i \to \tilde{\chi}^0_j Z$
are open and not suppressed, the $Z$ polarization reconstructed via
leptonic $Z$--boson decays with great precision allows us to probe
the Majorana nature and CP violation in the neutralino
system\cite{Twobody}.

\section{Implications for LC experiments of the SUSY DM scenario}
\label{sec:implications}

The DM constraints from the recent WMAP results\cite{WMAP} on the
SUSY parameter space imply, for many of the retained working points, a small
mass difference
$\Delta m \leq m/20$ between  the tau slepton
$\tilde{\tau}_1$, and the LSP mass for the so--called {\it co--annihilation
mechanism}. The amount of DM
depends critically on $m_{\tilde{\tau}_1}$ itself as well.
This means that {\it the proper justification of the co--annihilation mechanism
requires an extremely precise measurement of $m_{\tilde{\tau}_1}$ and
$m_{\tilde{\chi}^0_1}$.}

In this co--annihilation scenario the detection and the mass measurement
of the tau slepton through threshold scan is, however, challenging because
of a potentially very large background due to the four fermion final states,
the so--called $\gamma\gamma$ background.

A recent work in Ref.~\cite{SUSY_DM} has shown with a detailed analysis that
a forward veto to remove the $\gamma\gamma$ background down to
very small angles is essential to reach an almost background free result,
adequate to achieve the accuracy implied by the post--WMAP generation in
a model independent analysis. It has also pointed out the reduction
of efficiency of this veto by a non--zero crossing angle between electron
and positron beams and due to the very large overlaid background produced
by beam--beam interaction hitting the very forward electromagnetic calorimeter.

\section{NMSSM neutralino sector}
\label{sec:nmssm}

The NMSSM superpotential\cite{Miller:2003ay,Choi:2004zx} with an
iso--singlet Higgs superfield $\hat{S}$ in addition to the two
Higgs doublets superfields $\hat{H}_{u,d}$ reads
\begin{eqnarray}
W=W_Y +\lambda \hat{S}(\hat{H}_u
\hat{H}_d)+\frac{1}{3}\kappa\hat{S}^3 \label{eq:superpotential}
\end{eqnarray}
where $W_Y$ denotes the MSSM Yukawa components. The two
dimensionless parameters $\lambda$ and $\kappa$ are less than 0.7
with $\kappa < \lambda$ favored at the electroweak scale if
they remain weakly interacting up to the GUT scale\cite{Miller:2003ay}.

The singlet superfield adds an extra higgsino to the MSSM
neutralino spectrum, called a {\it singlino}, resulting in five
neutralinos. We denote the singlino dominated neutralino $\tilde
\chi_5^0$, with $\tilde \chi_{1-4}^0$ denoting the other four
neutralinos in order of ascending mass.

In the above preferred scenario, the singlino dominated neutralino  is the
lightest neutralino (and the LSP) with a mass of approximately
$\mu_{\kappa} \equiv 2 \kappa \langle S \rangle$ so that it will
be copiously produced at the LHC in squark and gluino cascade
decays. A very decoupled state with low $\lambda$ can give rise to
macroscopic flight distances of order a $\mu$m and order a nm for
the decays $\tilde \chi_1^0 \to \tilde \chi_5^0 l^+l^-$ and
$\tilde l_R \to \tilde \chi_5^0 l$ with
$\mu_{\lambda}(\equiv\lambda v/\sqrt{2})=1$~GeV, respectively. With the
integrated luminosity of $1 \; {\rm
ab}^{-1}$, large event rates of order $10^3$ are expected for
production of $\tilde{\chi}^0_5$, $\tilde{\chi}_1^0$ or $\tilde{\chi}_3^0$
with $\tilde{\chi}^0_5$ for $\mu_{\lambda} > 30$ GeV.
These characteristic signatures will allow us to distinguish the NMSSM
from the MSSM experimentally.

\section{Conclusions}
\label{sec:conclusions}

As evident from the examples discussed above and in a lot of collective
studies\cite{LC}, if a few sparticles are kinematically accessible, the LC
will enable us to make model--independent measurements of a host of
SUSY parameters and to reveal a
variety of phenomenological implications.

The highest possible precision to be provided by the LC ($\otimes$ LHC)
experiments\cite{LC_LHC} is essential to reveal the SUSY structure and breaking
mechanism through a reliable grand extrapolation to the Planck scale.
This should definitely be one of the most important aspects of the LC physics
potential.

\section*{Acknowledgments}

The author would like to thank his collaborators including
J. Kalinowski, Y.G. Kim, Y. Liao, J.S. Lee, D.J. Miller, G. Moortgat--Pick,
M. M\"{u}hlleitner, M. Spira and P.M. Zerwas for fruitful collaborations.

\end{document}